\documentclass[useAMS,usenatbib]{mn2e}
\usepackage{graphicx}

\title[Dynamical investigation of the Stothers idea]{Can turbulent convective variations drive the Blazhko cycle? Dynamical investigation of the Stothers idea}
\author[L. Moln\'ar, Z. Koll\'ath, and R. Szab\'o]{L. Moln\'ar\thanks{E-mail: lmolnar@konkoly.hu (ML);
kollath@konkoly.hu (ZK)}, Z. Koll\'ath, R. Szab\'o \\
Konkoly Observatory, MTA CSFK, Budapest, 1121, Konkoly Thege \'ut 13-17, Hungary}
\begin{document}

\date{2012 March 12.}

\pagerange{\pageref{firstpage}--\pageref{lastpage}} \pubyear{2010}

\maketitle

\label{firstpage}

\begin{abstract}
The Blazhko-phenomenon, the modulation of the pulsation of RR Lyrae stars remains one of the most stubborn unsolved problems of stellar pulsation. The recent idea of Stothers proposes that periodic variations in the properties of the convective envelope may be behind the amplitude and phase modulation. In this work we approximated the mechanism by introducing variations in the convective parameters of the Florida-Budapest hydrodynamic code and also by means of amplitude equations. We found that the process is only effective for long modulation periods, typically for more than hundred days, in agreement with the thermal time scales of the pulsation in RR Lyrae stars. Due to the slow response of the pulsation to the structure changes, short period, high amplitude Blazhko-modulation cannot be reproduced with this mechanism or would require implausible variations in the convective parameters on short time scales. We also found that the modulation of the mixing length results in strong differences between both the luminosity and radius variations and the respective phase modulations of the two quantities, suggesting notable differences between the energy output of the photosphere and the mechanical variations of the layers. The findings suggest that the convective cycle model is not well suited as a standalone mechanism behind the Blazhko-effect.
\end{abstract}

\begin{keywords}
stars: variables: RR Lyrae -- hydrodynamics -- convection
\end{keywords}

\section{Introduction}
The Blazhko-phenomenon, the amplitude and phase modulation of the pulsations of RR Lyrae stars remains one of the longest-lived mysteries in astrophysics. Despite of the elegant simplicity it displays at first sight, the Blazhko-effect seems to defy any theoretical suggestions. We summarise the observational knowledge and competing models here shortly, and refer to the excellent review of \citet{kovacs09} and references therein for further details. 

The Blazhko-effect was discovered about a century ago by \citet{blazhko} and \citet{shapley}. Since then, observations of RR Lyrae stars continued to accumulate, leading eventually to large photometric programs as the Konkoly Blazhko Survey led by Johanna Jurcsik or The Blazhko Project coordinated by Katrien Kolenberg, using dedicated telescopes (\citealt{jurcsik09}, \citealt{sodor07}), multi-site campaigns \citep{kolenberg06}, and intensive spectroscopic observations (\citealt{chadid06}, \citealt{kolenberg10sp}). Ground-based efforts uncovered numerous interesting details in the Blazhko variations of RR Lyrae stars: changing cycle length (\textit{e.g.} for RR Lyr  see \citealt{kolenberg06}), multiple period \citep*{sodor10} or irregularly disappearing modulation \citep{jurcsik09}. Longer variations as the famous ``4--year cycle'' in RR Lyr \citep{dsz73} were reported as well. 

The long and continuous observations of the CoRoT and \textit{Kepler} space telescopes have been providing further details. CoRoT uncovered a modulated RRab with variable Blazhko-amplitude \citep{gug11} as well as numerous Blazhko-sidelobes up to the 8th order in the Fourier-spectra of V1127 Aql \citep{chadid10}. The latter may indicate periodic modulation of the amplitude and the phase of the pulsation frequency (\citealt{szeidl09}, \citealt*{bszp11}). The \textit{Kepler} sample of more than 40 stars revealed many interesting cases, from highly variable modulation amplitudes to period doubling (\citealt{kolenberg10b}, \citealt{benko10}, \citealt{pd}, \citealt{gug12}). Additional frequencies close to the first and second radial overtone were also identified in both samples (\citealt{poretti10}, \citealt{benko10}). But from the theoretical side, instead of helping to single out a valid theory, all these nuances of Blazhko-variables emphasise the shortcomings of the current models.

\subsection{Classical Blazhko-models}
Until the appearance of the Stothers idea \citep{stothers06}, two models were considered as explanations. The magnetic oblique rotator model (\citealt{cousens}, \citealt{shibahashi00}) postulates a strong, $\sim 1$ kG dipole magnetic field inside the star, inclined to the rotational axis. The field distorts the radial pulsation, introducing an $l = 2$ spherical harmonic component, and the rotation of the star creates the modulation pattern, and a quintuplet structure in the frequency spectrum. There are two essential drawbacks of the magnetic oblique rotator model: no strong dipole fields were observed unambiguously in RR Lyrae stars (\citealt{chadid04}, \citealt{kolenbergmag}), and all the irregularities and complexities in the Blazhko-variation (as in the case of RY Com for example, \citealt{jurcsik08}) contradict with the simple geometric, rotation-based explanation.

The competing model has been the nonradial resonant rotator which ties the Blazhko modulation to a 1:1 resonance between the radial and a (preferably $l=1$, $m=1$) nonradial mode (\citealt{nowa}, \citealt{dziem}). Because of the current lack of nonradial, nonlinear pulsation models, the resonance model was developed using the amplitude equation (AE) formalism since AEs represent hydrodynamic calculations well. The nonradial resonant rotator model, similarly to the magnetic model, proposes a rotation-based mechanism and predicts symmetric modulation triplets. Observations are however, contradictory: beside the problems with a clockwork-like explanation mentioned above, recent, detailed studies show that side-peaks are usually neither symmetric nor limited to triplet components (see \textit{e.g.} \citealt{jurcsik05b}, \citealt{hurta08}, \citealt{sodor10b} and \citealt{chadid10}).

The most recent proposal \citep{bk11} was also based on AEs, incorporating the high-order radial resonance that was found to drive the period-doubling phenomenon \citep*{kmsz11}. The calculations revealed periodically and irregularly oscillating amplitudes that resemble the Blazhko-modulation quite well. These results suggest a more intimate relation between radial mode resonances and the Blazhko-effect.

\subsection{The idea of Stothers}
The proposed mechanism of \citet{stothers06}, although lacks detailed elaboration, seems to face a number of problems. The idea goes as follows: a turbulent magnetic field builds up in the outer layers of the star, stalling the convection. Then the field is destructed somehow, by ohmic decay or convective shredding for example. The periodic changes in the turbulent convective properties of the star then---in principle---are sufficient to change the amplitude and period of the pulsation, creating the observed Blazhko-effect. 

Unlike models that rely on stellar rotation, such mechanism with stochastic components would allow small changes and irregular variations in the modulation. Similarly, magnetic cycles would explain the variations over longer time scales, like the 4--year cycle of RR Lyr. A strong but turbulent magnetic field could explain the lack of detection as current observations are well suited only to dipole fields. But there is no mention in the original article \citep{stothers06} of the required field strength and configuration or how the pulsation itself would interact with the mechanism. \citet{kovacs09} discusses these shortcomings in more detail. The dynamics of the process are also totally omitted as radiative and convective single-mode models are compared to each other. It is quite possible that on time-scales of the shortest observed modulations (a few weeks), the stellar interior simply cannot change efficiently if at all to produce observable modulation. This was already implied by \citet{smolec11}, concluding that reproducing modulations, similar of observed in RR Lyr, requires huge modulation in the mixing length parameter. We are interested in a more general question: how does the observed variation depend on the modulation periods, \textit{i.e.} how effective is the response of the pulsation to the internal modulation of the properties of the convective layer? We will investigate these problems both with hydrodynamic calculations (Sections \ref{sectmodel} and \ref{sectnonlin}) and amplitude equations (Section \ref{sectampeq}).

\section{The model and the approach}
\label{sectmodel}
General investigation of the proposed mechanism would require full three-dimensional MHD modelling of the pulsating stars, including radial pulsation, convection and magnetic field evolution. Due to the current lack of such models, we explore the overall dynamics of such modulation with the Florida-Budapest code, a one-dimensional, turbulent convective hydrodynamic code. Treatment of the time-dependent turbulent convection is based on the method developed by \citet{kuhfuss86}. 

The code involves eight dimensionless $\alpha$ free parameters of which seven are independent from each other. They are connected to the turbulent convective properties of the model and are of order unity. Numerical values of $\alpha$ parameters are not provided by theory however, only the comparison with observations provide guidance. We can exploit this freedom to introduce modulations in the convective zone.

The best-known parameter is the mixing length ($\alpha_\lambda$ or $\alpha_{ML}$), and it is often used to fine-tune the convective properties to match the observations. \citet{smolec11} showed that large variations in the mixing length produce modulation and variations in the period-doubling phenomenon. However, we usually follow a different path by setting $\alpha_\lambda = 1.5$ as a constant value, and fine-tune the model with the other parameters. In those cases we do not consider the mixing length as an independent parameter.

A more detailed description of the model is given in \citet{kollath01} and \citet{kollath02}, and we also refer to \citet{kmsz11}, where we used the same model to investigate the period doubling in RR Lyrae stars. We also apply the amplitude equation method \citep{buchler84} to describe the modulation characteristics.

\begin{figure}
\includegraphics[width=85mm]{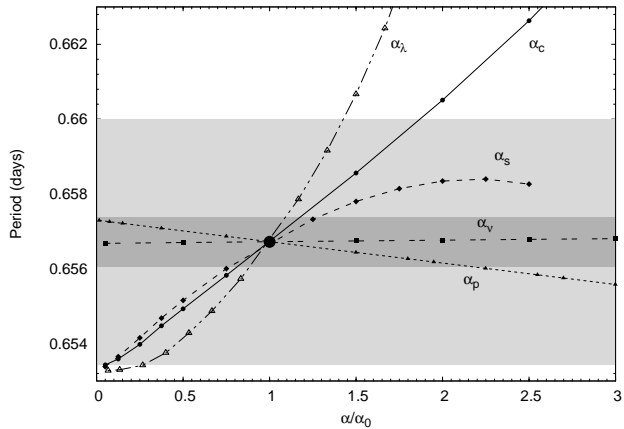}
\caption{Linear fundamental period values versus normalized $\alpha$ parameters. The large black dot denotes the reference values of all four parameters. Dark and light grey areas show the $\pm 0.1\%$ and $\pm0.5\%$ changes to the reference periods respectively. Variations of pulsation periods of those orders were found in modulated RR Lyrae stars.}
\label{linper} 
\end{figure}

\subsection{Linear results}
\label{linres}
Before running the time-consuming non-linear calculations to obtain amplitudes, we carried out a linear analysis. Here we summarise the results published in \citet{molnar10} shortly. 

Phase modulation, the change of pulsation period in Blazhko RR Lyrae stars usually does not exceed 1-2 $\%$. Quasi-continuous, space-based photometry allowed to calculate the instantaneous periods for a few modulated stars: four CoRoT RR Lyrae stars revealed variations ($\Delta P/P$ total amplitudes) between $0.2 - 1.3\%$ \citep{corot_pm} while V783 Cyg, observed with \textit{Kepler}, showed period change of $\sim 1.1\%$ \citep{kolenberg10b}. We estimated period changes from published phase modulation plots and pulsation-modulation frequency values in \citet{molnar10}: four stars, SS Cnc \citep{jurcsik06}, RR Lyr \citep{kolenberg06}, AR Her \citep{smith99} and MW Lyr \citep{jurcsik08} have quite similar values to each other and to V783 Cyg, between $0.8-1.2\%$ while for RR Gem \citep{jurcsik06} and DM Cyg \citep{jurcsik09a} it is somewhat lower, $\sim 0.3\%$. To achieve such changes in the linear periods however, quite large variations are required in convective parameters. Not all types of $\alpha$ parameters are even suitable: for example the eddy viscosity ($\alpha_\nu$) has very little effect on linear periods since it does not change the equilibrium stellar structure. The best candidates we found were the mixing length itself ($\alpha_\lambda$) and the parameters controlling the convective flux ($\alpha_c$) and the turbulent source function ($\alpha_s$), but even those would require huge modulation amplitudes (Figure \ref{linper}).

Changes in pulsation amplitudes are related to the growth rates of eigenmodes so we examined those dependencies as well. We approximate the relation from the amplitude equation method. The simplest form of an AE with a single mode present is $\dot A = \kappa A - q A^3$, where $A$ is the amplitude, $\kappa$ is the linear growth rate of the mode and $q$ is the saturation term respectively. By considering a limit cycle solution ($\dot A = 0$) and a constant $q$ saturation term, one can estimate the magnitudes of amplitude variations by $A \sim \sqrt\kappa$. By examining the various relations between the growth rates and the $\alpha$ parameters, we concluded that variations in $\alpha_c$, $\alpha_s$ and $\alpha_\nu$ may be suitable (see figure 2.\ in \citet{molnar10}). But even so, creation of high-amplitude modulation requires huge changes that cannot be justified on physical grounds in these convective parameters.

\section{Nonlinear model calculations}
\label{sectnonlin}
Our goal was to determine whether changes over the typical Blazhko-period time scales in the convective environment could create suitable modulation in amplitudes and phases. We do not attempt to find or validate any underlying physical processes behind the convective modulation. In that sense the modulation introduced in the model is just as \textit{ad hoc} as in original article of \citet{stothers06}: some parameters of the stellar structure are varied, and the response is observed. There is a crucial difference though: instead of comparing unmodulated models with different convective parameters, we perturb a limit-cycle solution with time-dependent, sinusoidal modulation and observe the variations it creates in the global parameters like radius or luminosity. We expect that the response of the convective envelope will strongly depend on the period of the modulation. 

\begin{table}
 \caption{Global parameters of the model and the $\alpha$ parameters of the convective zone (mean values for the modulated ones). }
 \label{param_table}
 \begin{tabular}{cccccccc}
  \hline
  \multicolumn{2}{c}{$T_{eff}$} & \multicolumn{2}{c}{$M$} 
    & \multicolumn{2}{c}{$L$} & \multicolumn{2}{c}{$Z$} \\
  \multicolumn{2}{c}{6100 $K$} & \multicolumn{2}{c}{0.77 $M_\odot$} 
    & \multicolumn{2}{c}{50 $L_\odot$} & \multicolumn{2}{c}{$10^{-4}$} \\
  \hline
  $\alpha_\nu$ & $\alpha_c$ & $\alpha_d$ & $\alpha_s$ & 
  $\alpha_t$ & $\alpha_p$ & $\alpha_r$ & $\alpha_\lambda$ \\
  0.2 & 0.2 & 8.0 & 0.2 &
  0.3 & 0.667 & 0.4 & 1.5 \\
  \hline
 \end{tabular}
\end{table}

The model in use is the same as in \citet{molnar10}, parameters are listed in Table \ref{param_table}. These values place it outside the period doubling instability region (see figure 5. in \citealt{kmsz11}). Models that reached the limit cycle were iterated with modulated $\alpha$ parameters. We considered the effects of the mixing length and some individual parameters as well. One reason behind this approach is that turbulent parameters do not necessarily change in accordance in a three-dimensional, interacting convective environment as they do in the one-dimensional, mixing length controlled models. We particularly emphasize the eddy viscosity parameter ($\alpha_\nu$) as it is strongly connected to the kinetic energy through the viscous dissipation and thus to the radius variations of the model. Mean radius variations during the Blazhko-cycle can be derived from multicolour photometry using the Inverse Photometric Method (\citealt{sjsz09}, \citealt{sodor09}). 

\begin{figure}
\includegraphics[width=85mm]{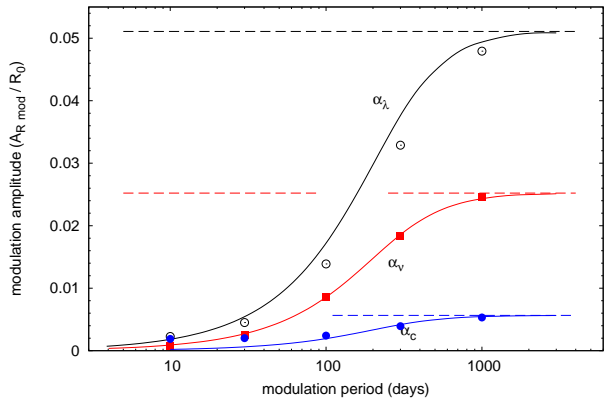}
\caption{Response to an internal modulation in radius variations as a function of modulation period. Squares, dots and cicles represent hydrodynamic calculations whereas red, blue and black lines represent the amplitude equation results for modulated $\alpha_\nu$, $\alpha_c$ and $\alpha_\lambda$ parameters, respectively. The dashed lines show the differences between models calculated with the two extreme values of the $\alpha$ parameters, \textit{i.e.} maximum values the modulation may reach (the $B(t)$ forcing amplitude introduced in Eq.\ \ref{eq2}).}
\label{alphan_ampeq} 
\end{figure}

\begin{figure}
\includegraphics[width=85mm]{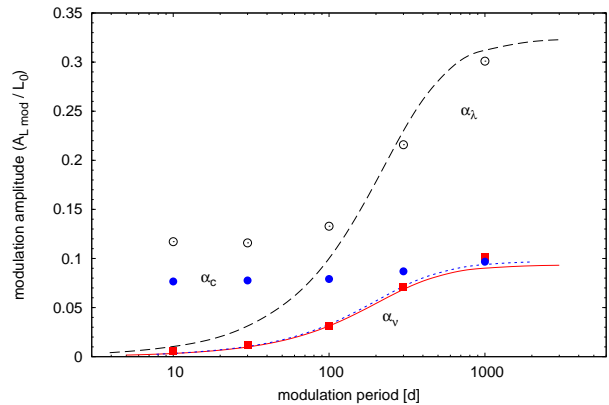}
\caption{Same as Figure \ref{alphan_ampeq} but for luminosity variations. Squares, dots and circles represent hydrodynamic calculations whereas red, blue dotted and black dashed lines represent the amplitude equation results for modulated $\alpha_\nu$, $\alpha_c$ and $\alpha_\lambda$ parameters, respectively. Note the difference between the hydrodynamic calculations and amplitude equation results at shorter periods.}
\label{lumin} 
\end{figure}

The modulation periods were 10, 30, 100, 300 and 1000 days, the modulation amplitude of $\alpha$ parameters was constant in all cases ($\pm 25 \%$). The results confirmed our expectations: the observed modulations in the radius and other global parameters grew with longer periods. The hydrodynamic model results are represented with various dots on Figures \ref{alphan_ampeq} and \ref{lumin}. Note that both the 10 and 30 day long modulation resulted in very low amplitude changes in most cases. The 10-day period value is close to the shortest modulation observed (SS Cnc: 5.31 d \citep{jurcsik06}, RR Gem: 7.23 d \citep{jurcsik05}) where amplitude changes are indeed relatively small: less than $0.1$ magnitudes in V colour and about $0.4 \%$ in mean radius \citep{sodor09}. The 30-day period is however in the range of many high-amplitude Blazhko-variables, including RR Lyr itself (38 d). Notable examples with short modulation periods but large amplitude variations include MW Lyr, where $P_m = 15.6$ d, and the mean radius variation is calculated to be $1.1\, \%$ \citep{jurcsik08} and the doubly-modulated star CZ Lac, with Blazhko-periods $P_{m1}\approx 18.6$ d, $P_{m2}\approx 14.5$ d, radius variations of 0.2 and $0.7\, \%$ and brightness variations of 0.3 magnitudes for both modulation components \citep{sodor10}. These values are quite similar to the results obtained for RZ Lyr, a star with significantly longer modulation components ($P_{m1}\approx 121$ d, $P_{m2}\approx 30$ d, \citealt{jurcsik12}). Yet the models show higher-amplitude modulation close to and above 100 days only, a distinction that does not apply for the Blazhko-stars. We note however that for a direct comparison, modulation amplitudes of the radius shall be determined along with the variations of the mean properties.

Not all convective parameters result in the same modulation properties. For the radius variations (Figure \ref{alphan_ampeq}), both the mixing length ($\alpha_\lambda$) and the eddy viscosity ($\alpha_\nu$) display a strong modulation period dependence. The convective flux parameter ($\alpha_c$), although it generates only marginal modulation, nevertheless shows some offset from the corresponding amplitude equation results (see Section \ref{sectampeq}). Differences are more pronounced in the case of luminosity (Figure \ref{lumin}), where the modulation amplitude corresponding to the $\alpha_c$ parameter is almost constant for all periods and it is notably higher, equivalent to the highest $\alpha_\nu$ values. This difference arises from the various effects of the modulations created in the model. Coefficients for the amplitude equations were determined from model series with constant convective parameters. If the convective parameters, especially $\alpha_c$ are modulated, the layer we perceive as the stellar photosphere also varies. In addition, the exact determination of the radius and the luminosity in the model is not straightforward and could result in differences between models with constant and dynamically changing convective parameters. This effect is also clearly visible for short periods if the mixing length is modulated, but for periods longer than about 150 days, the hydrodynamic and AE calculations agree: the modulation amplitudes increase with longer modulation periods.

Another feature that differs from the observations is the variation of pulsation maxima and minima. Blazhko variables show stronger modulation in maxima, our models however create more symmetrical variations or sometimes even stronger in the minima when transformed to bolometric magnitudes. Such discrepancy could be attributed however both to the nature of the internal modulation or to the limitations of the description of the stellar atmosphere of the model.

\subsection{Amplitude versus phase modulation}
An interesting aspect is the relation between the modulation of amplitude and phase in the variations of some global properties. Here we introduced sinusoidal modulations driven either by the three parameters and compared the properties of radius and luminosity variations. In the case of modulated $\alpha_\nu$, radius and luminosity behave similarly: the magnitude of modulation decreases towards shorter modulation periods. In other words, the luminosity modulation mostly reflects the mechanical variations of the layers of the model. 

\begin{figure}
\includegraphics[width=85mm]{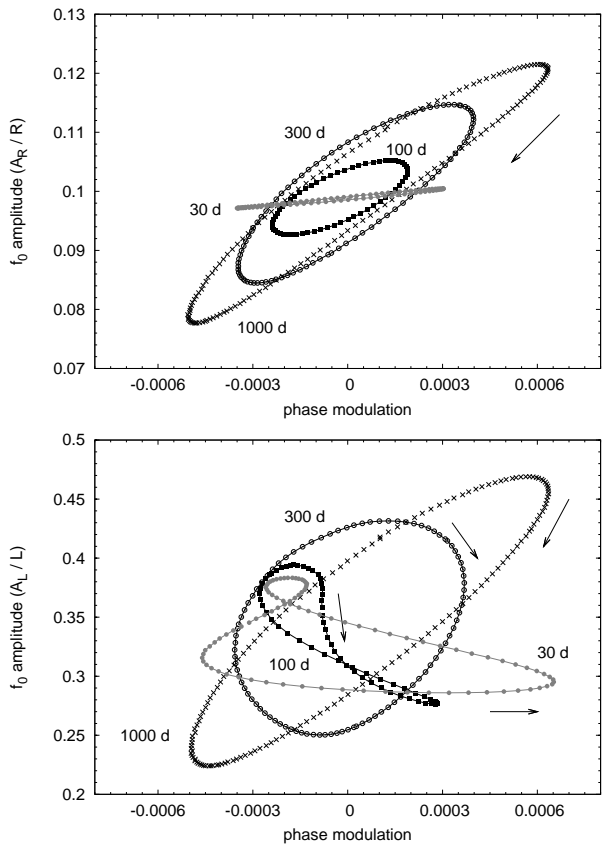}
\caption{Amplitude and phase modulation observed in the radius (above) and luminosity (below) variations at different modulation periods in the model. The large difference between the two properties suggest that the photosphere itself experiences strong variations if the mixing length ($\alpha_\lambda$) is varied. Arrows denote the direction of progression. Note the looping of the luminosity-phase curve at 100 and 30 days. }
\label{phasemod} 
\end{figure}

If we modulate the mixing length, the changes in radius and luminosity variations during the modulation cycle display striking differences. The radius variation is relatively simple, with decreasing amplitudes and simple curves. On the other hand, the phase of the luminosity variation changes more wildly and gets stronger towards short modulation periods and a looping of the phase relation curve is also visible. The modulation properties of radius and luminosity variations are displayed in Figure \ref{phasemod}. 

These calculations indicate that the modulation of the mixing length results in significant differences between the oscillating mechanical system and the energy output of the photosphere. The effect is strongest if we modulate the convective flux parameter ($\alpha_c$) only, but appears in the mixing length at shorter modulation periods as well. The most striking phenomenon is the looping of the luminosity-phase curves at 30 and 100 days modulation periods. This feature, if confirmed, could provide an additional test in itself for the convective cycle hypothesis. DM Cyg \citep{jurcsik09a} shows hints of looping but that feature might arise from the large scatter of points. Also, the modulation period is 10.57 days, and the model with 10-day modulation (omitted from Figure \ref{phasemod} for clarity) showed no looping. The direction of progression is also affected by the looping: long-period modulations progress in a clockwise manner but the the large loop dominating the 30-day curve and the entire 10-day curve is purely counter-clockwise. 

We emphasise however that the current description of the stellar atmosphere in the model is not well suited to investigate the properties of the photosphere in a dynamically changing environment. A more elaborate analysis with a more detailed atmospheric model will be required before the observed phase relations (luminosity, radius, pulsation phase) and model results can be compared directly.

\begin{figure}
\includegraphics[width=85mm]{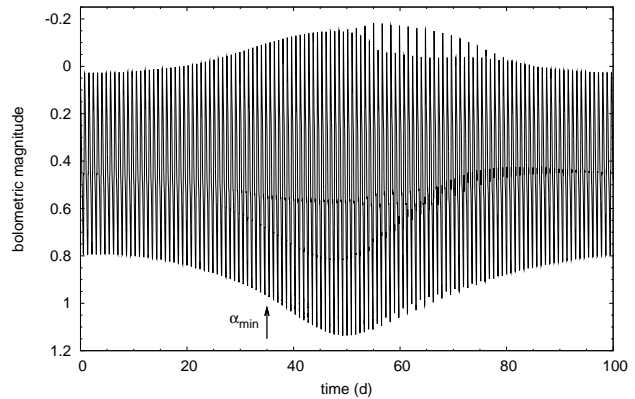}
\caption{Modulated nonlinear model with period doubling. Modulation period is 100 days, variations in the $\alpha_c$ and $\alpha_\nu$ are $\pm 75\%$. Period doubling occurs around and after the peak amplitude. The time of minimum $\alpha$ values is indicated with the arrow: maximum amplitude and period doubling occurs noticeably later. Model parameters are: $T_{eff}=6400\,K$, $L=60\,L_\odot$, $M=0.55\,M_\odot$. }
\label{nonlin-pd} 
\end{figure}

\subsection{Period doubling}
The continuous and precise data of the \textit{Kepler} space telescope revealed period doubling in some modulated RR Lyrae stars (\citealt{kolenberg10b}, \citealt{pd}), triggering extensive hydrodynamic modelling that traced the origin of the phenomenon back to a 9:2 resonance between the fundamental mode and the 9th radial overtone \citep{kmsz11}. Period doubling was also reproduced in the models of \citet{smolec11}. Although these results tend to favour the mode resonance and convective-cycle models of the modulation compared to the oblique rotator scenario, neither was able to reproduce the modulation itself (without introducing it artificially). On the other hand, \citet{bk11} incorporated the 9:2 resonance into amplitude equations and found not only period doubling (two-mode fixed points in the AE formalism) but modulated solutions as well.

By modulating models that fall into the PD instability region, the amplitude variation and/or appearance and disappearance of the PD phenomenon during the modulation cycle can be compared to the observations. On one hand, data from the first 127 days of \textit{Kepler} (Q1 and Q2) indicate that period doubling is the strongest on the descending or ascending branches of the Blazhko-modulation, and variations from one Blazhko-cycle to another are also obvious \citep{pd}. On the other hand, period doubling occurs predominantly during the Blazhko maxima in the models, as already pointed out by \citet{smolec11}, and the cycles are repetitive, as expected from a repetitive modulation. The inspection of Floquet stability roots \citep{kmsz11} also showed that instability is expected towards less efficient convection (smaller $\alpha$ parameters, closer to the radiative limit) where the pulsation amplitudes are higher. 

A nonlinear model with amplitude modulation and period doubling is shown in Figure \ref{nonlin-pd}. Parameters are: $0.55\, M_\odot$, $60\, L_\odot$, $6400 K$, modulated parameters are $\alpha_c$ and $\alpha_\nu$, both by $\pm 75 \%$ with a period of 100 days. Unusually, period doubling appears around 
the peak amplitude and dominates the descending branch of the amplitude modulation. The effect is very similar if the mixing length itself is modulated.

\section{Amplitude equation calculations}
\label{sectampeq}
Nonlinear model calculations are time-consuming exercises especially if long-period modulation is involved. If the time scales are separated enough, amplitude equations can be used instead of the hydrocode, excluding the pulsation from the integration and focusing only on the amplitude variations \citep{buchler84}. Coefficients have to be determined first from the models but the freedom of changing them is another possible advantage. The effects of different growth rates for example can be investigated much faster this way. 

As noted in Section \ref{linres}, the simplest amplitude equation with only one mode present is $\dot{A}=\kappa A- q A^3$. The coefficients of AEs were determined the following way. Linear growth rates of the fundamental mode ($\kappa(\alpha)$) were calculated in the linear models with different $\alpha$ parameters. Then the corresponding nonlinear models were iterated and saturation coefficients were determined from constant amplitude, limit cycle models $(A=A_0)$, in the form of $q(\alpha) = \kappa / A_0^2 $. We considered variations in three model parameters, $\alpha_{\nu}$, $\alpha_c$ and $\alpha_\lambda$. For all parameters, the equation: 
\begin{equation}
 \dot{A}(t) = \kappa(\alpha(t))\,A - q(\alpha(t))\,A^3
 \label{eq1}
\end{equation}
had to be solved. By setting $\kappa / q = B^2$, the equation can be expressed as:
\begin{equation}
\frac{\dot{A}}{A}=\kappa \left( 1 - \frac{A^2}{B^2} \right)
\label{eq2}
\end{equation}
where $B$ is the forcing amplitude exerted by the changes of the stellar interior: without the non-zero reaction time of the pulsation, the observable amplitude would be $B(t)$. Amplitude changes are however governed by the above equation which is a nonlinear filter with phase shift and a frequency response defined by $\kappa$.
To create large variations over shorter time scales, $B$ has to be similar or even higher than $A$ which is physically implausible. These properties are shown by the integration of (\ref{eq1}) in the following subsections.

\subsection{Response to an internal modulation}
We calculated AEs with a constant modulation amplitude ($\alpha_\nu$, $\alpha_c$ and $\alpha_\lambda$ varied by $\pm25\%$) and a wide range of modulation periods. The observed modulation in the stellar radius is plotted in Figure \ref{alphan_ampeq}. The solid lines show the AE results while hydrodynamic calculations are represented with various dots for comparison. Amplitude equations represent hydrodynamic calculations well in general, however some differences arise in our case. As we explained it in Section \ref{sectnonlin}, these effects are caused by the inherent differences between the calculated luminosity and radius values of constant and modulated hydrodynamic models. Even with these restrictions, the results clearly show that short and even medium-period modulation ($<100$ days) results in low-amplitude modulation of the observable quantities. The highest possible amplitudes (\textit{i.e.} the difference between the two extreme states, dashed lines) are reached only at very long modulation periods, over 1000 days. The values represented by the dashed lines are similar to the $\delta R/ R$ value used by \citet{stothers11}, the difference between the radial and convective models. 

In all three cases the observed modulation amplitudes increase linearly from short to long modulation periods before approaching the maximum value asymptotically. The only difference is in the highest value: modulation of the convective flux has the smallest effect on the radius variations followed the eddy viscosity and the mixing length. The luminosity values are equally affected by the former two parameters but both create about the third of the effect of mixing length only. Modulating the system with larger amplitudes allow for larger observed variations on shorter time scales as well but in that case even stronger modulation should occur over long periods. Of course the possibility of slow but very strong modulation cannot be ruled out by the lack of observations. A much stronger argument against such mechanism is the existence of fast, strong modulation in some RR Lyrae stars: it would require enormous changes in the convective properties of the stellar envelope over only a matter of weeks.

\begin{figure}
\includegraphics[width=85mm]{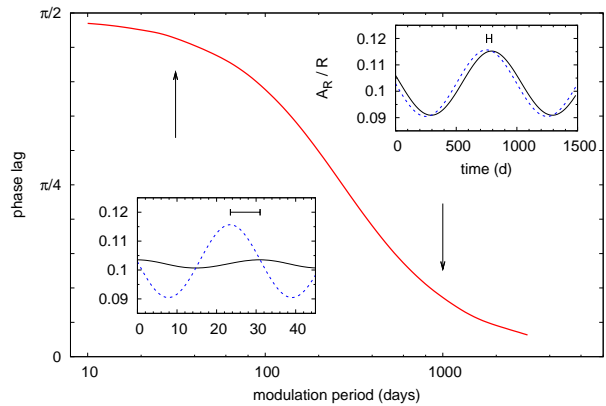}
\caption{Phase difference between the modulation of the $\alpha_\nu$ parameter and the radius variations versus the period of the modulation. Values approach $\pi/2$ for moderate and fast modulation. Inserts show two cases: the modulation period for the upper right panel is $P=1000d$ while for the lower left panel $P= 31d$. Blue dashed lines show the $B(t)$ forcing amplitudes while the solid black lines show the $A(t)$ response amplitudes. The phase differences between the maxima of the two variations are indicated with the black bars. }
\label{phaselag} 
\end{figure}

\subsection{Phase relations}
The slow response of the pulsation to the internal modulation also manifests itself as a phase shift between the internal modulation cycle and the observable variation of the pulsation amplitude. The phase difference is shown in the main plot of Figure \ref{phaselag}. For modulation periods under about hundred days the difference approaches $\pi/2$. Two examples are shown in the subplots with modulation periods $P=1000$ and $P=31$ days. Solid and dashed lines represent the $A(t)$ response (observable) and $B(t)$ forcing amplitudes as defined in (\ref{eq2}). The decrease of amplitudes with shorter modulation periods is also very prominent. These results are confirmed with the corresponding hydrodynamic calculations as well. Figure \ref{nonlin-pd} shows that both the highest-amplitude pulsation and period doubling occur after the $\alpha$ parameters reached their minimum values.

\subsection{Changing the growth rate of the mode}
\begin{figure}
\includegraphics[width=85mm]{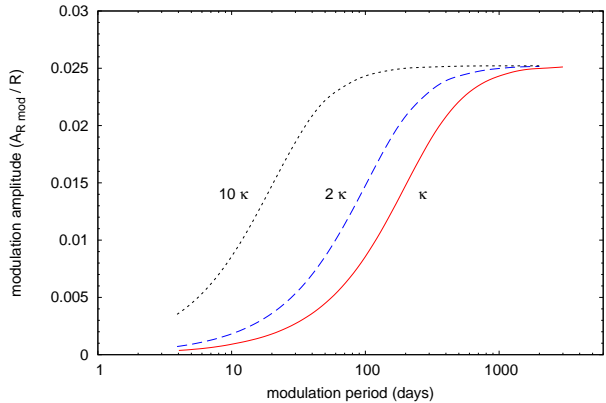}
\caption{Changing the growth rate of the mode shifts the curve along the time axis. Higher growth rates would allow faster modulation as well. The red solid line ($\kappa$) is the same as the red solid line ($\alpha_\nu$) on Figure \ref{alphan_ampeq}. The blue dashed ($2\kappa$) and black dotted ($10\kappa$) lines show the effect of higher growth rates.}
\label{alphan_kappa_q} 
\end{figure}

The normalised linear growth rate of the fundamental mode in RR Lyrae models is typically in the range of one per cent. Amplitude equations allow us to change these values easily and compare cases with different growth rates. Figure \ref{alphan_kappa_q} shows the effect of multiplying the growth rates by 2 and 10. Since $A_0^2=\kappa/q$ in our simple model, the saturation term was multiplied as well to fix the magnitude of the amplitudes. Varying these parameters essentially scales the time parameter to an arbitrary unit. But these simple exercises further confirm that the slow response of the pulsation mode is the main drawback for the Stothers-model: the $10 \kappa$ (tenfold increase in the growth rate) case allows much faster amplitude changes: large amplitude variations over 10-20 pulsation periods, or a few dozen days. In contrast, \citet{bk11} showed that resonant coupling with the 9th overtone can generate adequate amplitude modulation with reasonable mode growth rates.

\section{Conclusions}
In this paper we investigated some dynamical aspects of the idea proposed by \citet{stothers06} to explain the Blazhko-effect. Although the advocated scenario is very complicated, the idea is based on the simple comparison of some properties (amplitudes, periods) of radiative and convective RR Lyrae models. The properties of the mechanism, involving a highly variable turbulent magnetic field coupled with the convective properties of the envelope, represent great challenges both for observations and detailed stellar models. We restricted ourselves to a quite simplified method by planting an \textit{ad hoc} internal modulation into a 1D turbulent convective model. A similar approach was followed by \citet{smolec11} to recreate the modulation properties of RR Lyr. However, instead of focusing on the best reproduction of the modulation in a single star, we carried out a more general analysis, studying the possible modulation amplitudes over a broad range of modulation periods. 

As the mean pulsation period also changes during the Blazhko-cycle, we first compared the periods of linear models with different convective parameters. The two suitable parameters to achieve the required variations ($\le 1\%$) were the ones controlling the convective flux and the turbulent source function but even those require huge variations. These results already present restrictions for the further, non-linear calculations. 

Blazhko RR Lyrae stars come in all flavours, most importantly in all kinds of different modulation periods and amplitudes. The greatest challenges for the mechanism are the ones with strong amplitude variations over short time scales, under about 40--50 days. We investigated the dynamics of the mechanism by modulating various convective parameters (eddy viscosity, strength of the convective flux and mixing length) in the model and by comparing the observed amplitude variations of the global stellar parameters over different modulation periods. We found that the mechanism is efficient only for long modulation periods ($>100$ days) but fails to explain the short-period, large-amplitude Blazhko stars. Increasing the internal modulation also postulates huge amplitude changes for very long modulation periods. These results are in agreement with the findings of \citet{smolec11}. It is worth mentioning that the changes applied to convective parameters in this paper or by \citet{smolec11} are still less than those in the original paper \citep{stothers06} where simply a fully convective and a fully radiative model were compared. 

The reason of the ineffectiveness of the mechanism is the slow response of the pulsation to the changes in the stellar envelope. Typical normalised growth rates of the modes in RR Lyrae stars are in the range of $\sim 10^{-3}-10^{-2}$. The mechanism of Stothers would require modes with an order of magnitude faster growth rates to allow efficient amplitude changes over a few dozen pulsation cycles or less. The results also suggest that although the difference between the convective and radiative models seem to account for the observed mean radius changes \citep{stothers11}, the process is dynamically incapable to reproduce the variations under the modulation periods mentioned in the paper (\textit{e.g.}\ 15.6 days for MW Lyr). The actual difference between the two models is reached only when incorporating years-long modulation periods.

The results also indicate that the luminosity and thus the properties of the photosphere may exhibit variations which are different from the radius that reflects the changes in the pulsation energy only. We found that the observed luminosity modulation is less dependent on the modulation periods but even in the best case, only moderate modulation could arise for modulation periods under about 150 days. We note however that a more detailed description of the stellar atmosphere is needed to draw further conclusions (\textit{e.g.} phase relations) about the properties of the photosphere in a dynamically changing environment.

All these results suggest that the mechanism proposed by Stothers faces at least as serious shortcomings as any of the current Blazhko-models. Most prominently, it is not suitable to explain strong modulation on short time scales especially without involving large amounts of internal variations in the convective parameters. Our analysis of course neglects a number of processes, \textit{e.g.} the feedback of pulsation to the magnetic field and the dynamical coupling between the convective envelope and the variable magnetic field is only crudely approximated with the internal modulation. Without such a detailed modelling effort, the plausibility of the model cannot be rejected. Instead of a standalone mechanism though, it may be incorporated into a more complex explanation of the Blazhko-effect along with the recent findings about period doubling and radial mode resonances.

\section*{Acknowledgments}

This work was supported by the Hungarian OTKA grants K83790, K81421 and MB08C 81013. RSz acknowledges the Bolyai J\'anos Scholarship of the Hungarian Academy of Sciences.

\bsp

\label{lastpage}

\end{document}